\documentclass{sadhana}

\usepackage{graphicx,amsmath,bm}
\usepackage{xcolor}
\usepackage{nomencl}
\usepackage{tabularx}
\usepackage{setspace}
\newlength{\bibitemsep}
\newlength{\bibparskip}
\let\oldthebibliography\thebibliography
\renewcommand\thebibliography[1]{%
	\oldthebibliography{#1}%
	\setlength{\parskip}{\bibitemsep}%
	\setlength{\itemsep}{\bibparskip}%
}
\makenomenclature
\begin{document}

\title{Relative accuracy  of turbulence simulations using pseudo-spectral and finite difference solvers}


\author{Akash Rodhiya\textsuperscript{1}, Shashwat Bhattacharya\textsuperscript{2,*}\and Mahendra K Verma\textsuperscript{1}}
\affilOne{\textsuperscript{1} Department of Physics, Indian Institute of Technology Kanpur, Kalyanpur, Kanpur 208016, India\\}
\affilTwo{\textsuperscript{2} School of Mechanical and Materials Engineering, Indian Institute of Technology Mandi, Kamand, Mandi 175005, India\\ e-mail: shashwat@iitmandi.ac.in}


\twocolumn[{

\maketitle

\begin{abstract}
For a single timestep, a spectral solver is known to be more accurate than its finite-difference counterpart. However, as we show in this paper, turbulence simulations using the two methods have nearly the same accuracy.  In this paper, we simulate forced hydrodynamic turbulence on a uniform 256$^3$ grid for Reynolds numbers 965, 1231, 1515, and 1994. We show that the two methods yield nearly the same evolution for the total energy and the flow profiles. In addition, the steady-state energy spectrum, energy flux, and probability distribution functions of the velocity and its derivatives are very similar. We argue that within a turbulence attractor, the numerical errors are likely to get cancelled (rather than get added up), which leads to similar results for the finite-difference and spectral methods. These findings are very valuable, considering that a parallel finite-difference simulation is more versatile and efficient (for large grids) than its spectral counterpart. 
\end{abstract}

\msinfo{2 December 2024}{5 August 2025}{13 August 2025}


}]


\setcounter{page}{1}
\corres
\volnum{123}
\issuenum{4}
\monthyear{January 2016}
\pgfirst{2333}
\pglast{2335}
\doinum{12.3456/s78910-011-012-3}
\articleType{}


\markboth{Rodhiya, Bhattacharya, and Verma}{Relative accuracy of pseudo-spectral and finite difference solvers}

\section{Introduction}
\label{sec:level1}
There are several numerical approaches for solving fluid flow. In this paper, we will compare and contrast two important and popular approaches -- the pseudo-spectral method and the finite-difference method. In the finite-difference (FD) method, the governing equations are solved by approximating the derivatives using finite differences. The errors in FDs are minimized by reducing the distance between two successive gridpoints or by increasing the order of the FD method~\cite{Ferziger:book:CFD,Anderson:book:CFD}. 
For example, \textcolor{black}{the error in computing the first derivative using the central difference scheme is of the order $O(\Delta x)^2$}, where $\Delta x$ is the distance between two successive grid points. 
In a pseudo-spectral method, the spatial derivatives are calculated using Fourier, Chebyshev, or Legendre transforms. 
The derivative of a function computed using the pseudo-spectral method is exact as long as the function is fully resolved~~\cite{Boyd:book:Spectral,Canuto:book:SpectralFluid}.
An argument based on spectral analysis shows that for the same accuracy, the required number grid points in a pseudo-spectral method is approximately six times less than that for the FD method~\cite{Hildebrand:book:Numerical, Verma:book:Python}.
Due to these reasons, pseudo-spectral codes are widely used for accurate turbulence simulations. {\color{black}In periodic domains, the pseudo-spectral method has been employed to simulate fluid turbulence with Reynolds numbers (based on the large-scale eddies) ranging from  $O(10^2)$~\cite{Orszag:PRL1972} to as large as  $O( 10^5)$~\cite{Yeung:CPC2025}. Pseudo-spectral solvers are also used} in weather forecasting and in instability studies~\cite{Boyd:book:Spectral,Canuto:book:SpectralFluid,Chandrasekhar:book:Instability}. 

A major limitation of pseudo-spectral solvers is that they are difficult to implement for non-periodic and complex boundaries. Further, pseudo-spectral codes have a low parallel efficiency due to MPI \texttt{Alltoall} communications among all the processors. The FD methods, on the other hand, are more amenable to parallelization~\cite{Ferziger:book:CFD} and can simulate flows in complex geometries. Hence, FD methods are being used increasingly to simulate turbulence in domains with non-peri-odic boundary conditions (see, for example, Verzicco and Camussi \cite{Verzicco:JFM2003}, Emran and Schumacher~\cite{Emran:JFM2008}, Akhmedagaev \textit{et al}~\cite{Akhmedagaev:JFM2020}, Gruber \textit{et al}~\cite{Gruber:CF2021}, and Pandey \textit{et al}~\cite{Pandey:JFM2022}).  
\textcolor{black}{However, for domains with periodic boundaries, the FD methods have not been used as extensively as pseudo-spectral methods;  only a few such studies exist for a limited range of Reynolds numbers from $O(10^2)$~\cite{Samtaney:PF2001} to $O(10^3)$~\cite{Larkermani:CMAME2024}. 
	This is because} for the same resolution, the error per timestep is higher for a FD solver than that of its pseudo-spectral counterpart. \textcolor{black}{In spite of the above}, as we will show in this paper, turbulence evolution by the two methods are quite similar \textcolor{black}{for the same grid size}.

There are only a handful of works that compare the accuracies of the FD and pseudo-spectral methods.
Fornberg~\cite{Fornberg:Geopphys1987} simulated elastic wave equation using the two methods and showed that for the same grid, the spectral method provides more accurate solutions than the FD solver, but it takes a longer computer time.
Recently, 
Verma~\textit{et al}\cite{Verma:SNC2020} carried out a comparative performance analysis between a FD solver and a spectral solver for simulating \textcolor{black}{decaying} turbulence at Reynolds number 1000 and reported the two codes to have comparable accuracy.
In this paper, we carry out a more comprehensive comparison between the FD and pseudo-spectral methods for \textcolor{black}{simulating forced turbulence} at higher Reynolds numbers (up to $\mathrm{Re} \approx 2000$) and show that the results of both the solvers are similar. 

Turbulence simulations with high Reynolds numbers require fine grids along with large memory and computing time, which is possible only on parallel high-performance computing systems~\cite{Yokokawa:CP2002,Yeung:PNAS2015,Yang:IEEE2016,Krasnov:JCP2023}. Most of these simulations employ many compute processors. However, in recent times the usage of graphics processing units (GPUs) for accelerating fluid flow solvers has gained interest in the scientific community~\cite{Thibault:CP2009,Czechowski:CP2012,Niksiar:IEEE2014,Ravikumar:SC2019,Piscaglia:Aerospace2023,Zehner:CF2023,Verma:SNC2023}. This is due to GPU's higher computing capabilities and less power consumption compared to CPUs~\cite{Owens:IEEE2008,Li:NCA2019}, along with the availability of programming platforms such as CUDA and CuPy~\cite{Luebke:IEEE2008,Sanders:book:CUDA,Strontium:book:CUDA,Ansorge:book:CUDA,CUDA:Web,Nishino:CP2017,Cupy:Web}. The simulations presented in this paper are GPU-accelerated on CuPy.  
\begin{table*} [t!]
	\centering
	\caption{\label{tab:simDetails}Details of the three-dimensional direct numerical simulations using Py-Tarang and Py-Saras.  On NVIDIA RTX 3090 GPU, each timestep of Py-Saras and Py-Tarang takes approximately 1.5  and 0.8 seconds respectively. }
	\begin{tabular}{ |c|c|c|c|c| } 
		\hline
		Simulation & grid size & viscosity $\nu$ & Reynolds number & $k_\mathrm{max}\eta$ \\
		\hline
		Run A & 256$^3$ & 0.01 & 965 & 3.3\\ 
		Run B & 256$^3$ & 0.008 & 1231 & 2.8\\ 
		Run C & 256$^3$ & 0.0065 & 1515 & 2.4\\ 
		Run D & 256$^3$ & 0.005 & 1994 & 1.9\\ 
		\hline
	\end{tabular}
\end{table*}

\textcolor{black}{The main objective of this paper is to study the relative accuracies of turbulence simulations using FD and pseudo-spectral methods for Reynolds numbers ranging from 965 to 1994. We compare the total kinetic energy, vertical vorticity structures, and the probability distribution of the velocity fields along with their derivatives computed using the two solvers. Additionally, we examine and contrast the spectral results produced by the two solvers. 
	We also compare the speed of the two solvers in simulating the above flow in terms of wall time per time step after GPU acceleration.
}

The outline of the paper is as follows. In Section~\ref{sec2:num_solve}, we introduce the pseudo-spectral and FD solvers, following which we detail the cases that are examined. 
In Section~\ref{sec3:results}, we contrast the results generated by the solvers. We conclude in Section~\ref{sec4:conclusion}.

\section{Numerical Method}
\label{sec2:num_solve}
In this paper, we solve the incompressible Navier--Stokes equations using pseudo-spectral and FD methods. The Navier Stokes equations are
\begin{eqnarray}
	\frac{\partial \mathbf{u}}{\partial t}+ \mathbf{u} \cdot \nabla \mathbf{u} &=& -\frac{\nabla p}{\color{black}\rho} + \nu \nabla^2 \mathbf{u}, + \mathbf{F},\label{eq:Momentum} \\
	\nabla \cdot \mathbf{u} = 0, \label{eq:continuity}
\end{eqnarray}
where, $\mathbf{u}$ and $p$ are the velocity and pressure fields, respectively; $\nu$ is the kinematic viscosity; {\color{black}$\rho$ is the density;} $t$ denotes time; and $\mathbf{F}$ is the body force applied on the fluid. The body forces could be gravitational force, buoyancy force, or Lorentz force. In this paper, we employ Taylor-Green force~\cite{Cichowlas:PRL2005}.

The FD solver, Py-Saras, is the Python version of C++ Saras~\cite{Samuel:JOSS2020} which solves Eq.~(\ref{eq:Momentum}) and (\ref{eq:continuity}) for Cartesian grids. Py-Saras is written using an object-oriented structure. It employs the Marker and Cell method (MAC)~\cite{Harlow:PF1965} for discretizing the velocity and pressure fields.
It uses a second-order central difference scheme for computing the first and second derivatives. 
The user has the option of employing either the semi-implicit Crank-Nicolson method~\cite{Crank:PCPS1947} or the low-storage Runge-Kutta method~\cite{Williamson:JCP1980} (RK3) for time advancement. 
The continuity equation [Eq.~(\ref{eq:continuity})] is satisfied by solving a Poisson equation for a pressure correction field that is related to an intermediate velocity field~\cite{Gholami:SIAM2016}.
The Poisson equation is solved using the geometric multigrid method~\cite{Briggs:book:Multigrid,Wesseling:book:Multigrid} employing the Full Multigrid (FMG) V-cycle.

The pseudo-spectral solver, Py-Tarang (Python version of C++ Tarang~\cite{Verma:Pramana2013tarang}), solves the Navier Stokes equations in the Fourier space as follows:
\begin{eqnarray}
	\frac{d}{dt}\hat{u_i}(\mathbf{k},t) &=& -\sqrt{-1} \, k_i \, \frac{\hat{p}(\mathbf{k},t)}{\color{black}{\rho}} \nonumber \\    
	&&-\sqrt{-1} \, \sum_j k_j \, \widehat{u_j u_i}    
	-\nu \, k^2 \, \hat{u}_i (\mathbf{k}) + \hat{F}_i, 
	\label{eq:Momentum_Fourier} \\
	k_i \, \hat{u}_i (\mathbf{k},t) &=& 0,
\end{eqnarray}
where $\mathbf{k}$ is the wavenumber vector; and $\hat{p}$, $\hat{u}_i$, and $\hat{F}_i$ are, respectively, the Fourier transforms of the pressure field and  $i$th components of velocity and force fields.
The nonlinear term, $- \sqrt{-1} \, \sum_j k_j \, \widehat{u_j u_i}$, is computed using fast Fourier transform (FFT) to avoid convolutions in spectral space that are computationally expensive to calculate~\cite{Boyd:book:Spectral}. In this scheme, the fields are transformed from the Fourier space to real space, multiplied with each other, and then transformed back to the Fourier space. 
The Fourier modes can be advanced in time using Euler or Runge-Kutta schemes. Similar to Py-Saras, Py-Tarang is written in Python using an object-oriented structure.

We perform our turbulence simulations on a single NVIDIA RTX 3090 GPU.  We employ CuPy~\cite{Nishino:CP2017,Cupy:Web}, an open-source Python library that automatically converts the Python code to CUDA. CuPy uses CUDA-related libraries to make full use of the GPU's architecture.   Most array operations are vectorized. The acceleration process involves copying all the field data (stored as multidimensional arrays) from the CPU to the GPU and dividing the tasks among different GPU threads.

Using Py-Tarang and Py-Saras, we simulate four different cases of homogeneous isotropic turbulence in a cubical domain.  
\textcolor{black}{Although Py-Saras and Py-Tarang solve the dimensional form of the governing equations, we assume all the quantities to be dimensionless in the present work. Hence, the velocity, pressure, viscosity, density, domain length, and all other related quantities have non-dimensional units.}
The details of the cases with {\color{black}their respective parameters in their dimensionless forms} are mentioned in Table~\ref{tab:simDetails}.

\begin{figure}[t!]
	\centering
	\includegraphics[width=3.5in]{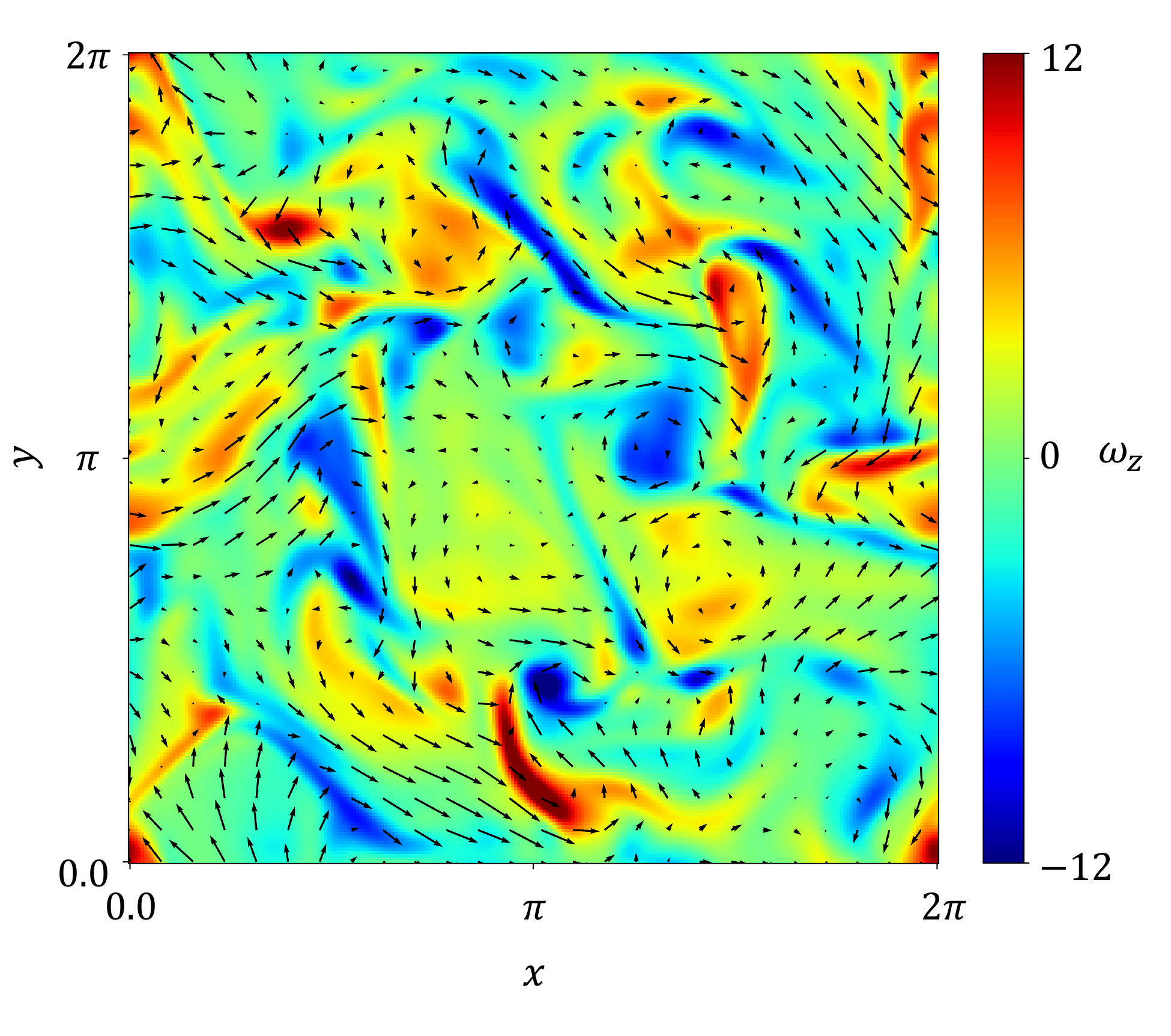}
	\caption{\label{fig1:init_cond} For the initial flow field for all the runs, the density plot of the \textcolor{black}{non-dimensional} vertical vorticity ($\omega_z$) superposed on the vector plot of the horizontal velocity field on the horizontal midplane.}
\end{figure}

All simulations are conducted on a domain of dimensions $2\pi \times 2\pi \times 2\pi$ using $256^3$ gridpoints. We impose periodic boundary conditions on all the walls. 
All the numerical simulations are provided with the same initial condition, which is a statistically stationary state of a homogeneous isotropic turbulence simulation. Figure~\ref{fig1:init_cond} illustrates the density plots of the vertical vorticity field ($\omega_z$) and the horizontal velocity vector plot at the horizontal midplane. The root-mean-square velocity (defined as $u_\mathrm{rms} = \sqrt{\langle u_x^2+u_y^2+u_z^2\rangle_V}$) of the initial flow field is 1.23 {\color{black}units}; here  $\langle \cdot \rangle_V$ represents a volume average. 

To sustain turbulence, we employ Taylor-Green force $\mathbf{F}$, which at wavenumber $k_0=1$ is given by:
\begin{equation}
	\mathbf{F}
	=
	F_0
	\begin{bmatrix}
		\sin(2\pi k_0 x)\cos(2\pi k_0 y)\cos(2\pi k_0 z) \\
		-\cos(2\pi k_0 x)\sin(2\pi k_0 y)\cos(2\pi k_0 z) \\
		0
	\end{bmatrix},
	\label{eg:forcing}
\end{equation}
where $F_0$ is unity.
The kinematic viscosities ($\nu$) for our runs are listed in Table 1.
We perform our simulations till 100 nondimensional time units using the third-order low-storage Runge-Kutta (RK3) scheme using a constant time step $dt = 0.001$ time units, where one time unit corresponds to $\sqrt{L/(2 \pi F_0)}$, where $L$ is the domain length.
Each time step of Py-Tarang and Py-Saras takes approximately 0.8 seconds and 1.5 seconds, respectively, on NVIDIA RTX 3090 GPU.
Thus, Py-Saras is nearly two times slower than Py-Tarang. This is because Py-Saras uses the Multigrid method for solving the pressure Poisson equation, which involves a series of restriction and prolongation operations.
On the other hand, Py-Tarang directly computes the pressure, and it hence is faster than Py-Saras.
However, on a distributed systems with many GPUs, we expect Py-Saras to be faster than Py-Tarang because \texttt{Alltoall} communications in FFTs incur huge costs compared to nearest-neighbor communication in the FD method.
\begin{figure}[b!]
	\includegraphics[width=8.7cm]{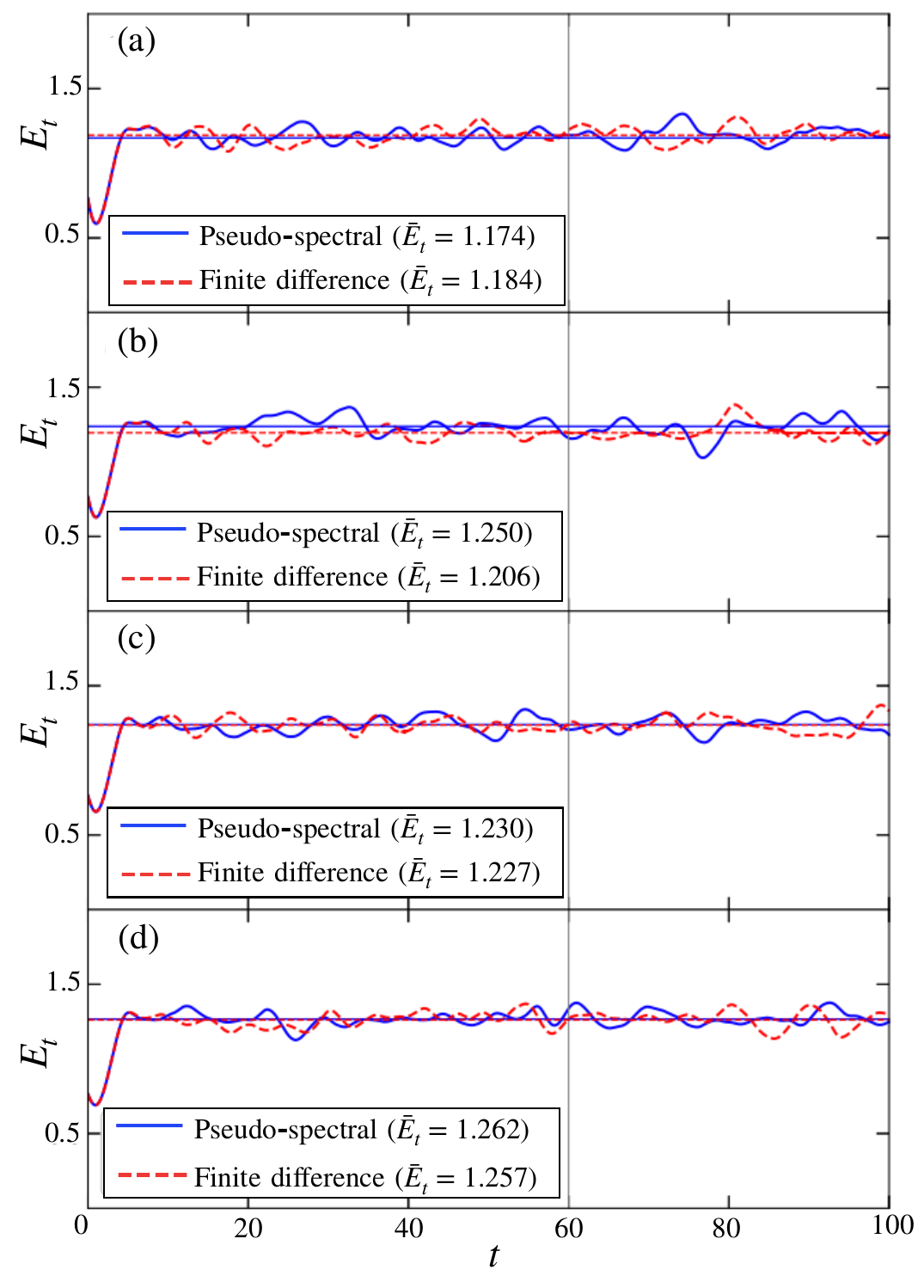}
	\caption{\label{fig:tot_ene} Comparative plots of the time series of the total energy ($E_t$) for (a) $\mathrm{Re}=965$, (b) $\mathrm{Re}=1231$, (c) $\mathrm{Re}=1515$, and (d) $\mathrm{Re}=1994$. {\color{black}The time-averaged values of the total kinetic energy ($\bar{E}_t$) are stated in the legends of each subplot and also denoted by a solid blue horizontal line for the pseudo-spectral solver and a dashed pink horizontal line for the FD solver.} The time series for the spectral and FD methods are similar. }
\end{figure}

All the runs reach their respective steady states after 10 time-units. In figure~\ref{fig:tot_ene} we plot the total energy ($E_t= u_\mathrm{rms}^2/2$) as a function of time $t$. The evolution of $E_t$ will be further explored in Sec.~\ref{sec3:results}.

\textcolor{black}{Table~\ref{tab:simDetails} also lists the Reynolds numbers (Re) for each run based on the mean of the time-averaged root mean square velocity computed by the two solvers at steady state.} The Reynolds number, which is defined as
\begin{equation}
	\mathrm{Re} = \frac{u_\mathrm{rms} \, L}{\nu},
\end{equation}
varies from 965 to 1994. For all our runs, the products of the largest wavenumber $k_\mathrm{max}$ and the Kolmogorov length scale $\eta$  are greater than 1.9 (see Table 1). Hence, our runs are well resolved~\cite{Boyd:book:Spectral}. Here, the Kolmogorov length scale is given by 
\begin{equation}
	\eta = \left( \frac{\nu^3}{\epsilon_u} \right)^{1/4},
\end{equation}
where $\epsilon_u$ is the viscous dissipation rate.

In the next section, we compare the results computed using both the solvers.

\section{Results}
\label{sec3:results}
{\color{black}In the following, we analyze and contrast the results obtained using Py-Saras (FD solver) and Py-Tarang (pseudo-spectral solver).} We focus our attention on the evolution of total energy, velocity and vorticity fields, energy spectra and fluxes, and the probability distribution functions of the velocity field and its spatial derivatives.

\subsection{Energy, velocity fields, and their derivatives}
\label{sec:RealSpace}

We compute the total energy ($E_t$) at each timestep using the numerical data generated by Py-Tarang  and Py-Saras. In figure~\ref{fig:tot_ene}, we plot the time series of $E_t$ for all the cases.
The figure shows that  $E_t$'s converge to statistically stationary values after $t=10$ and exhibit temporal fluctuations about their mean values. The mean values of $E_t$'s, shown as horizontal lines in figure~\ref{fig:tot_ene},  are computed by averaging the numerical data from $t=15$ to $t=100$. \textcolor{black}{The mean $E_t$'s of Py-Saras and Py-Tarang differ by $0.85\%$, $3.5\%$, $0.23\%$, and $0.37\%$,  for $\mathrm{Re}=965$, $1231$, $1515$, and $1994$, respectively.  Interestingly, the deviations appear to decrease with increasing Reynolds numbers. An explanation for this trend is provided later in this section.} 

The aforementioned results are rather surprising because we 
expect small errors in the velocity field to get amplified in a turbulent flow (like chaotic systems~\cite{Strogatz:book}). Hence, they 
appear contradictory to the assumption that for the same grid resolution, the errors generated by an FD code are more than those by a spectral solver. 
A possible reason for this result is that both methods lead the flow to the same attractor, and the errors during the evolution of the flow get canceled in both methods. 
In Sec.~\ref{sec4:conclusion} we discuss this issue in detail.
\textcolor{black}{Note that the time-scale of the fluctuations in lower Reynolds number flows is expected to be longer than those with larger Reynolds numbers~\cite{Frisch:book,Lesieur:book:Turbulence}. Hence, lower Reynolds number flows may require averaging over more timeframes for all the errors to get cancelled out. This might explain the trend of the deviations between the mean kinetic energy computed by Py-Saras and Py-Tarang in which the deviations appear to decrease with increasing Reynolds numbers.}

\begin{figure*}
	\centering
	\includegraphics[width=5.5in]{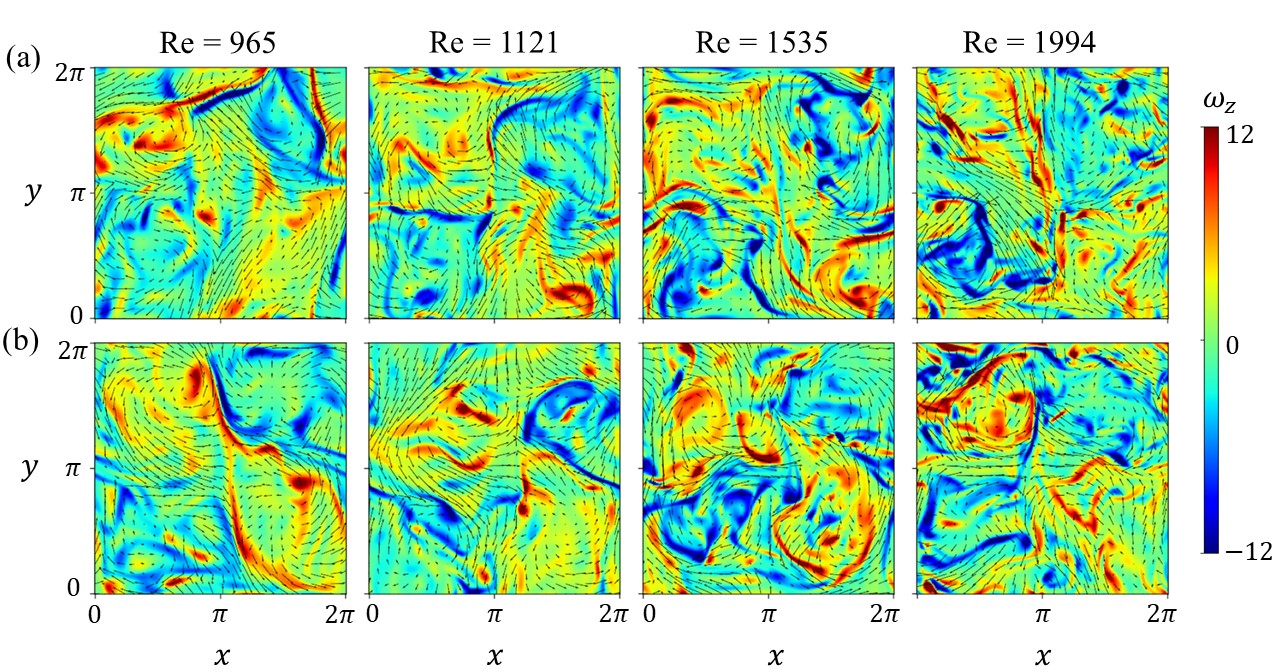}
	\caption{\label{fig3:contours_steady} Density plots of the \textcolor{black}{non-dimensional} vertical vorticity along with the vector plots of horizontal velocity vectors on horizontal midplane for different Reynolds numbers at $t=60$ for (a) pseudo spectral solver, and (b) finite-difference solver. \textcolor{black}{Although the structures are not exactly the same for the two solvers because of the turbulent nature of the flow, they have similar features in terms of size and density, and are statistically comparable.}}
\end{figure*}

\begin{figure*}
	\centering
	\includegraphics[width=5.5in]{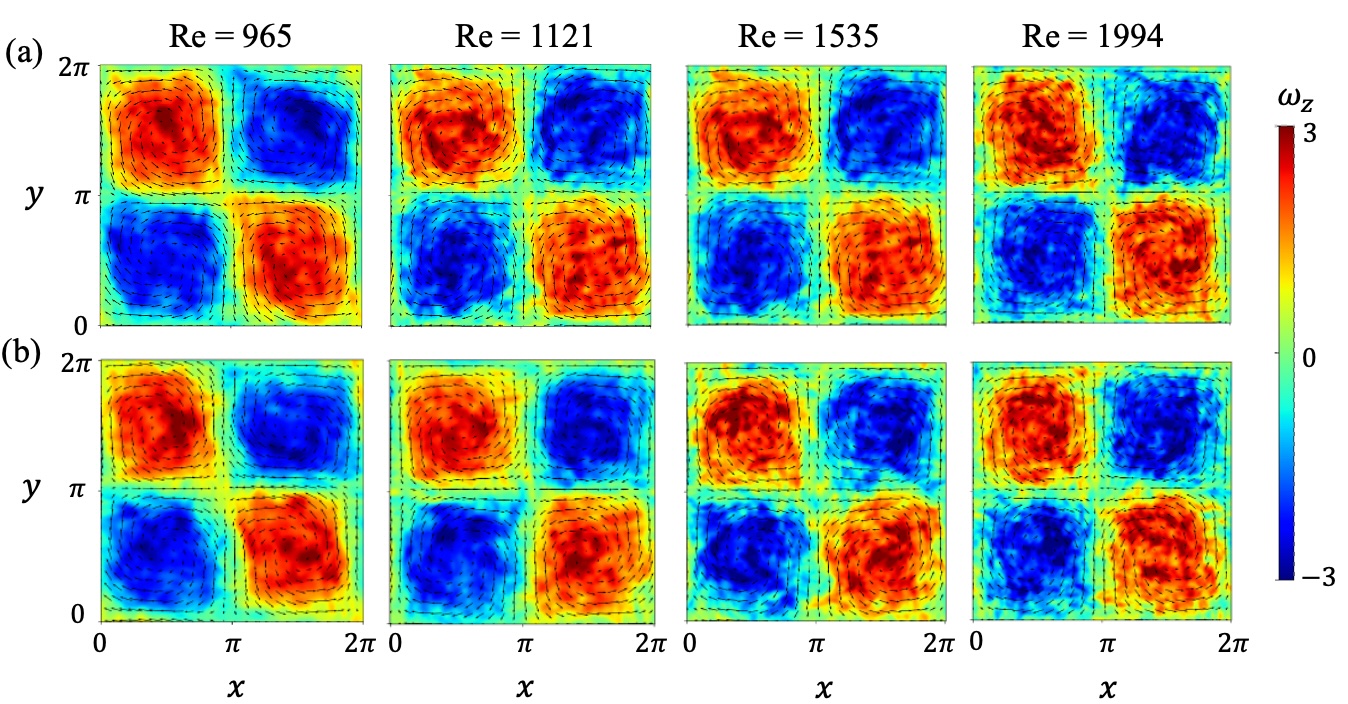}
	\caption{\label{fig:contours_timeAvg} \textcolor{black}{Density plots of the time-averaged non-dimensional vertical vorticity along with the vector plots of horizontal velocity vectors on horizontal midplane for different Reynolds numbers at $t=60$ for (a) pseudo spectral solver, and (b) finite-difference solver. The vorticity fluctuations get cancelled out due to time-averaging, and only the steady-state vortices generated due to the Taylor-Green force $\mathbf{F}$ remain.}}
\end{figure*}

For all the four runs,  in figure~\ref{fig3:contours_steady} we exhibit the velocity vector plots and the $z$-vorticity density plots on the the horizontal midplane at $t=60$ \textcolor{black}{In figure~\ref{fig:contours_timeAvg}, we exhibit the time-averaged plots of the aforementioned quantities.} Clearly, the \textcolor{black}{instantaneous} flow structures generated by the FD and spectral codes are not the same because of the turbulent nature of the flow.  However, the structures in the flow are statistically similar. For example, the rms value of $z$-vorticity for the two solvers deviates by only $2.5\%$. Also, the density of vortices increases with the increase of Re for both methods. 
Hence, the turbulence evolution by the two methods appears to be similar statistically, but not in detail. \textcolor{black}{On the other hand, the time-averaged plots shown in figure~\ref{fig:contours_timeAvg} appear similar because all the turbulent vorticity fluctuations get cancelled out due to averaging and only the steady-state vortices generated due to the Taylor-Green force $\mathbf{F}$ remain.} 

\begin{figure*}
	\centering
	\includegraphics[width=12.5cm]{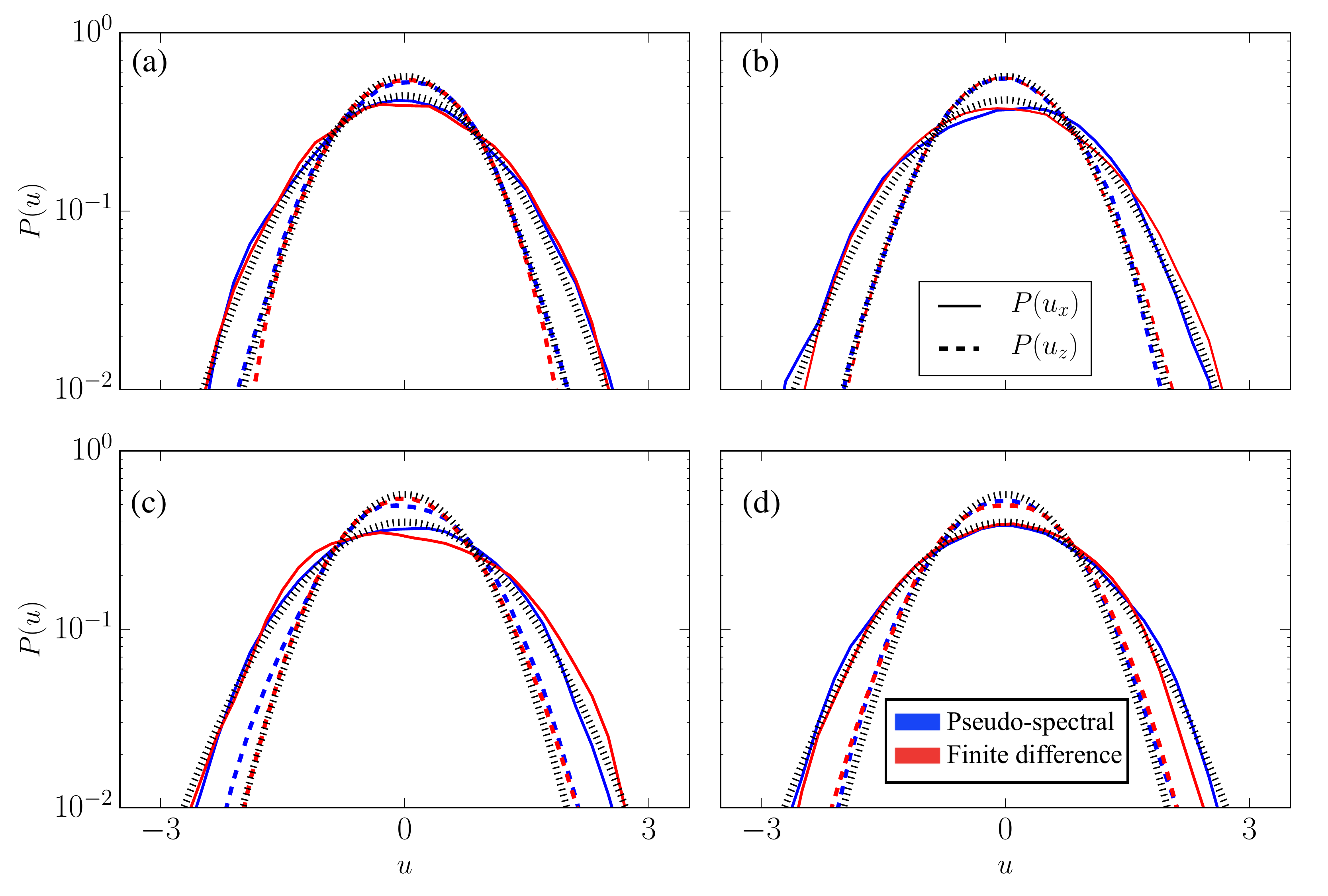}
	\caption{\label{fig3:vel_disti} Probability distribution functions of $u_x$ (solid curves) and $u_z$ (dashed curves) computed using the data of pseudo-spectral (blue curves) and FD (red curves) solvers for (a) $\mathrm{Re}=965$, (b) $\mathrm{Re}=1231$, (c) $\mathrm{Re}=1515$, and (d) $\mathrm{Re}=1994$. The PDFs closely follow the Gaussian distribution (black dotted curves) with the mean $\mu=0$ and the standard deviation $\sigma=0.7$ for $u_z$ and $0.9 \leq \sigma \leq 1$ for $u_x$.}
	\label{fig:PDF_u}
\end{figure*}

\begin{figure*}
	\centering
	\includegraphics[width=12.5cm]{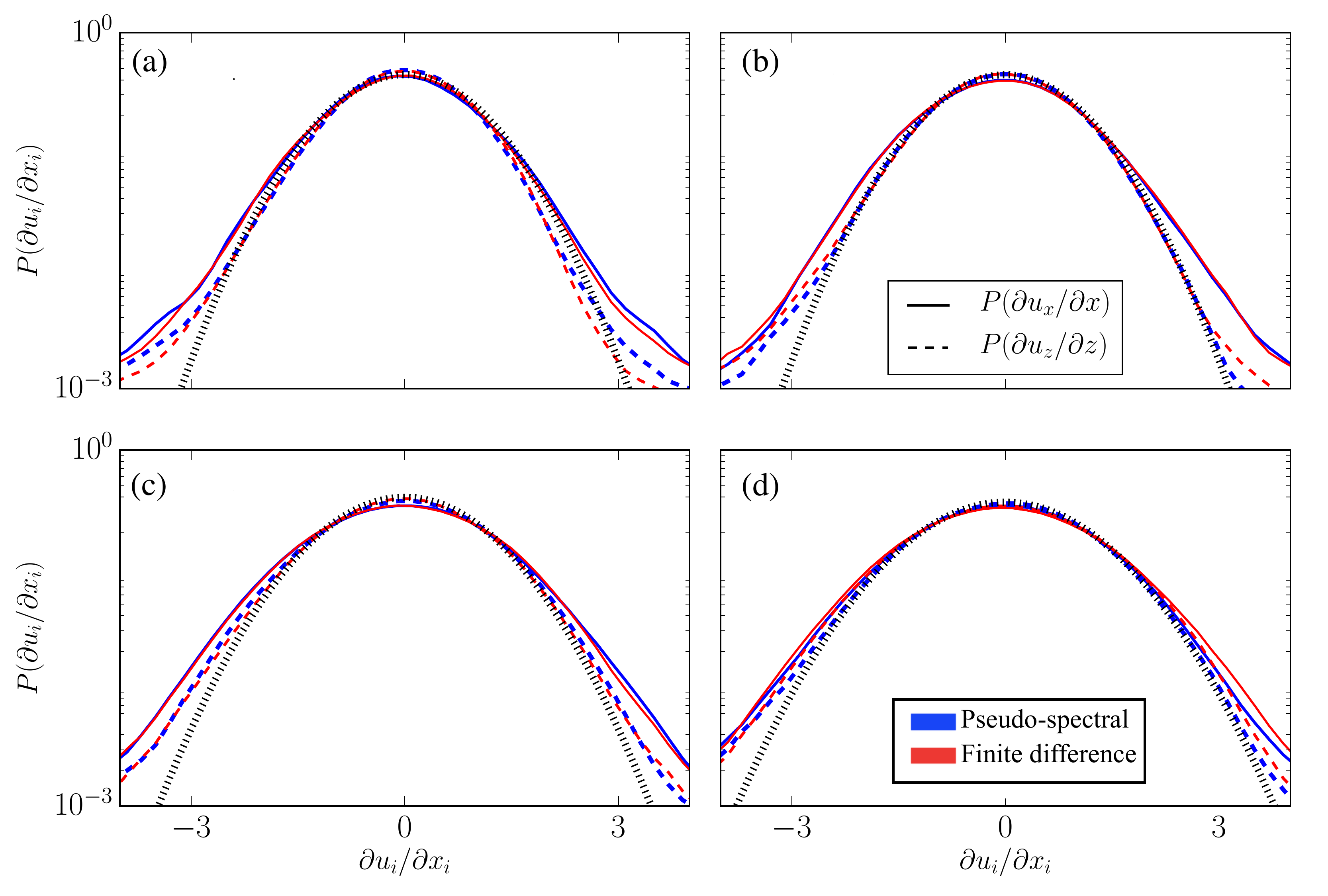}
	\caption{\label{fig3:veld_disti} Probability distribution functions of $\partial u_x/ \partial x$ (solid curves) and $\partial u_z/\partial z$ (dashed curves) computed using the data of pseudo-spectral (blue curves) and FD (red curves) solvers for (a) $\mathrm{Re}=965$, (b) $\mathrm{Re}=1231$, (c) $\mathrm{Re}=1515$, and (d) $\mathrm{Re}=1994$. The tails of the PDFs deviate from the Gaussian distribution (black dotted curves).}
	\label{fig:PDF_u_der}
\end{figure*}

{
	\begin{figure*}
		\centering
		\includegraphics[width=12.5cm]{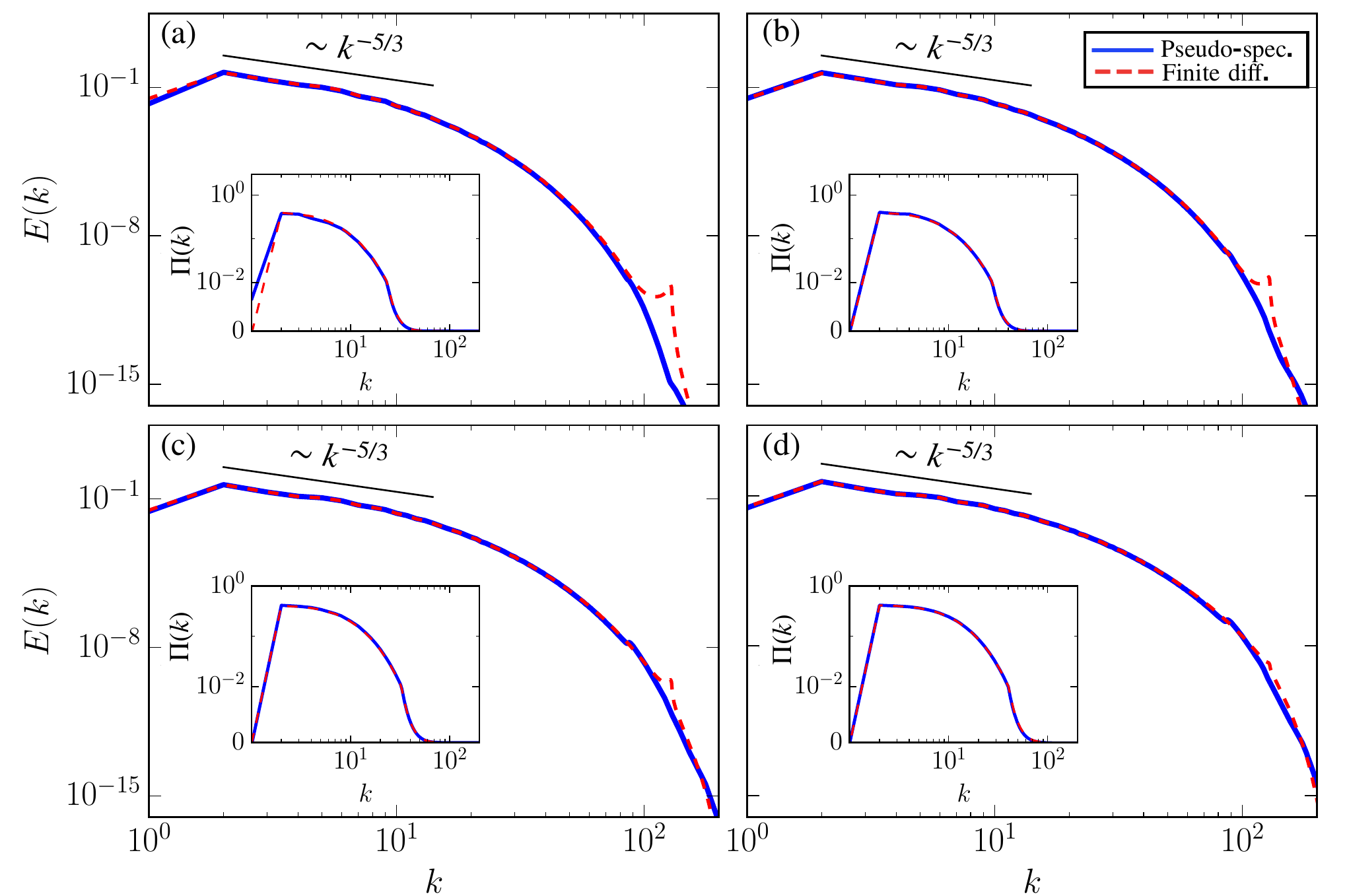}
		\caption{\label{fig3:spectra} \textcolor{black}{Energy spectra $E(k)$ computed using the data of pseudo-spectral (solid blue curves) and FD (dashed red curves) solvers for (a) $\mathrm{Re}=965$, (b) $\mathrm{Re}=1231$, (c) $\mathrm{Re}=1515$, and (d) $\mathrm{Re}=1994$. The inset in each subplot exhibits the respective energy fluxes.}}
		\label{fig:Ek}
	\end{figure*}

	We compute the turbulence statistics using the data generated by the two solvers. For the four runs, we calculate the probability distribution functions (PDF) of the horizontal velocity component $u_x$, the vertical velocity component $u_z$, and their respective longitudinal derivatives  $\partial u_x/\partial x$ and $\partial u_z/\partial z$, and plot them in figures~\ref{fig:PDF_u} and \ref{fig:PDF_u_der}.  
	The exhibited PDFs show that the aforementioned quantities for the two solvers are similar, except for the minor deviations in the tails of the PDFs (which may be due to dissimilar extreme events). As expected, the PDFs of the velocity field closely follows Gaussian distribution~\cite{Pope:book,Leslie:book}, as depicted by the best-fit dotted black curve of figure~\ref{fig3:vel_disti}. 
	
	The mean of the Gaussian distribution is zero for both velocity components, but the standard deviation for $u_z$ ($\sigma \approx 0.7$) is less than for $u_x$ ($\sigma \approx 0.9$). This is consistent with the extended tails of $u_x$ compared to $u_z$.  Unlike the velocity field, the PDFs of the velocity gradients deviate from the Gaussian distribution near the tails. Note that the deviations from the Gaussian behavior are stronger for  $\partial u_x/\partial x$ than $\partial u_z/\partial z$. These differences between $u_x$ and $u_z$ indicate anisotropy in the flows, which is due to the directional asymmetry of the  Taylor Green force. Our results are consistent with earlier studies~\cite{Vincent:JFM1991,Hwang:EF2004,Wilczek:JFM2011}. 
	
	\subsection{Spectral quantities}
	
	In Sec.~\ref{sec:RealSpace}, we showed that Py-Tarang and Py-Saras produce similar results in real space, generating closely matched total energy, PDFs of the velocity field, and qualitatively similar flow structures. In this subsection,  we will contrast the spectral results produced by the FD and pseudo-spectral solvers.  Note that the spectral quantities provide multiscale measures~\cite{Boyd:book:Spectral}.  Spectral solvers provide accurate spatial derivatives. Hence, we expect accurate turbulence simulations resolved all the way up to the smallest scales. In comparison, FD methods have discretization errors in the derivative computations.

	In this paper, we analyze two spectral quantities: the energy spectrum and the energy flux. The energy spectrum $E(k)$ is the kinetic energy contained in a wavenumber shell of radius $k$ and is given by 
	\begin{equation}
		E(k) =  \sum_{k\leq |\mathbf{k}'| < k+1} \frac{1}{2} |\hat{\mathbf{u}}(\mathbf{k}')|^2,  \label{eq:KEspectrum}
	\end{equation}
	where $\hat{\mathbf{u}}(\mathbf{k}')$ is the Fourier Transform of the velocity field. The energy injected at large scales cascades to smaller scales due to the nonlinear interactions among the velocity modes. This energy transfer is quantified by energy flux $\Pi_u(k_0)$, which is the kinetic energy leaving a wavenumber sphere of radius $k_0$. The flux is computed using~\cite{Kraichnan:JFM1959,Dar:PD2001}:
	\begin{equation}
		\Pi(k) = -\int_0^{k} dk' T_u(k'),
		\label{eq:flux_MtoM} \end{equation}
	where
	\begin{equation}
		T(k') = \mathbb{R}[ {\bf N}_u ({\bf k'}) \cdot {\bf u(k')}],
	\end{equation}
	with ${\bf N}_u ({\bf k})$ as the nonlinear term of the Navier-Stokes equation. 
	
	For all four cases, we compute the energy spectra using the velocity fields at $t=60$  and plot them in figure~\ref{fig3:spectra}. The energy flux is plotted in the insets of the same figure. Interestingly, $E(k)$ and $\Pi(k)$ reported by the two methods are nearly the same, except for the differences at large wavenumber ($k \gtrapprox 100$).  
	The $E(k)$'s computed using the FD  solver have kinks at large wavenumbers, which is due to the accumulation of energy at small scales and the dissipative nature of FD solvers.  For a narrow range of wavenumbers (called \textit{inertial range}),  $E(k) \sim k^{-5/3}$ and $\Pi(k) \approx \mathrm{const.}$, which is consistent with Kolmogorov's theory of turbulence~\cite{Kolmogorov:DANS1941Dissipation,Pope:book}. We expect a larger inertial range with higher-resolution simulations. 
	
	Thus, the two methods yield very similar $E(k) $ and $\Pi(k)$, implying that the energy contents, as well as the energy transfers, at all the scales, are nearly the same. This is rather surprising because the FD method is less accurate than the spectral method. We will discuss the possible reasons for these similarities in the next section.
	
	\section{Discussions and Conclusions}
	\label{sec4:conclusion}
	
	In this paper, we test the relative accuracies of turbulence simulations using FD and psuedospectral methods. For the same, we performed four turbulence simulations with Reynolds numbers 965, 1231, 1515, and 1994 on a $256^3$ grid using the same initial condition (a turbulent flow) and Taylor-Green large-scale forcing.  We observed that the two methods yield nearly the same total energy, energy spectra, and energy fluxes. The flow structures produced by the two methods are statistically similar, but they are not identical due to the growth of numerical errors with time.

	According to numerical analysis, the numerical error per timestep in the FD method is more than that in the spectral method. For example, for the same numerical accuracy, the FD scheme requires approximately 6 times more grid points compared to a spectral method~\cite{Hildebrand:book:Numerical,Verma:book:Python}.  Clearly, the results presented in the paper appear to be inconsistent with the above observations. We propose the following resolution to this puzzle. 
	
	At a given point of time, a turbulent flow profile is represented by a point in the phase space formed by the Fourier modes of the flow~\cite{Lesieur:book:Turbulence}. At the steady state, a turbulent flow evolves within a fixed phase space volume, called \textit{turbulent attractor}~\cite{Frisch:book,Lesieur:book:Turbulence,Orszag:CP1973}. Therefore, we expect the turbulence evolution by both the methods, FD and spectral, to be on this attractor.  Since the turbulence trajectories on the attractor are confined, the numerical errors do not add up during the evolution, but they possibly cancel each other so that the evolution remains inside the attractor. We expect that the spectral method will follow the \textit{real trajectory} (one without any numerical error) more closely than the FD method, but both methods evolve the trajectories in the attractor. This is the reason why both FD and spectral methods yield similar total energy, energy spectrum and flux, and PDFs. The above conjecture is akin to chaos theory's \textit{shadowing lemma}~\cite{Palmer:book:Chaos}, according to which a numerically computed trajectory of a nonlinear dynamical system follows a real trajectory quite closely. We plan to test our conjecture using more simulations.
	
	Our results encourage employment of FD method for accurate turbulence simulations. This is particularly useful for very large grids, for which spectral simulations are expensive due to the \texttt{Alltoall} communications involved in FFT. Also, FD methods can simulate flows with rigid walls, which is not easy to implement in spectral methods.

	\section*{Acknowledgement}
	The authors thank Soumyadeep Chatterjee, Roshan Samuel, Manthan Verma, and Mohammad Anas for useful discussions and development of the solvers. MKV acknowledges the support by Science and Engineering Research Board, India, for the Grants SERB/PHY/20215225, SERB/PHY/2021473, and J. C. Bose Fellowship (SERB /PHY/2023488). \\
	

\begin{thebibliography}{10}
	
	\bibitem{Ferziger:book:CFD}
	Ferziger J~H and Peric M
	\newblock 2001
	\newblock {\em {Computational Methods for Fluid Dynamics}}.
	\newblock Springer-Verlag, Berlin Heidelberg, 3 edition
	
	\bibitem{Anderson:book:CFD}
	Anderson J~D
	\newblock 1995
	\newblock {\em {Computational Fluid Dynamics: The Basics With Applications}}.
	\newblock McGraw-Hill, New York
	
	\bibitem{Boyd:book:Spectral}
	Boyd J~P
	\newblock 2003
	\newblock {\em Chebyshev and Fourier Spectral Methods}.
	\newblock Dover Publications, New York, 2nd revised edition
	
	\bibitem{Canuto:book:SpectralFluid}
	Canuto C, Hussaini M~Y, Quarteroni A, and Zang T~A
	\newblock 1988
	\newblock {\em {Spectral Methods in Fluid Dynamics}}.
	\newblock Springer-Verlag, Berlin Heidelberg
	
	\bibitem{Hildebrand:book:Numerical}
	Hildebrand F~B
	\newblock 2003
	\newblock {\em {Introduction to Numerical Analysis}}.
	\newblock Dover Publications Inc, New York, 2nd edition
	
	\bibitem{Verma:book:Python}
	Verma M~K
	\newblock 2021
	\newblock {\em {Practical Numerical Computing Using Python: Scientific and
			Engineering Applications}}.
	\newblock Self-published, Kindle Direct Publishing, 2nd edition
	
	\bibitem{Orszag:PRL1972}
	Orszag S~A and {Patterson, Jr} G~S
	\newblock 1972
	\newblock {Numerical Simulation of Three-Dimensional Homogeneous Isotropic
		Turbulence}.
	\newblock {\em Phys. Rev. Lett.}, 28:76
	
	\bibitem{Yeung:CPC2025}
	Yeung P~K, Ravikumar K, Nichols S, and Uma-Vaideswaran R
	\newblock 2025
	\newblock {GPU-enabled extreme-scale turbulence simulations: Fourier
		pseudo-spectral algorithms at the exascale using OpenMP offloading}.
	\newblock {\em Comput. Phys. Commun.}, 306:109364
	
	\bibitem{Chandrasekhar:book:Instability}
	Chandrasekhar S
	\newblock 1981
	\newblock {\em {Hydrodynamic and Hydromagnetic Stability}}.
	\newblock Dover publications, Oxford
	
	\bibitem{Verzicco:JFM2003}
	Verzicco R and Camussi R
	\newblock 2003
	\newblock {Numerical experiments on strongly turbulent thermal convection in a
		slender cylindrical cell}.
	\newblock {\em J. Fluid Mech.}, 477:19--49.
	
	\bibitem{Emran:JFM2008}
	Emran M~S and Schumacher J
	\newblock 2008
	\newblock {Fine-scale statistics of temperature and its derivatives in
		convective turbulence}.
	\newblock {\em J. Fluid Mech.}, 611:13--34
	
	\bibitem{Akhmedagaev:JFM2020}
	Akhmedagaev R, Zikanov O, Krasnov D, and Schumacher J
	\newblock 2020
	\newblock {Turbulent Rayleigh-B{\'e}nard convection in a strong vertical
		magnetic field}.
	\newblock {\em J. Fluid Mech.}, 895:R4
	
	\bibitem{Gruber:CF2021}
	Gruber A, Bothien M~R, Ciani A, Aditya K, Chen J~H, and Williams F~A
	\newblock 2021
	\newblock {Direct Numerical Simulation of hydrogen combustion at auto-ignitive
		conditions: Ignition, stability and turbulent reaction-front velocity}.
	\newblock {\em Combust. Flame}, 229:111385
	
	\bibitem{Pandey:JFM2022}
	Pandey A, Krasnov D, Sreenivasan K~R, and Schumacher J
	\newblock 2022
	\newblock {Convective mesoscale turbulence at very low Prandtl numbers}.
	\newblock {\em J. Fluid Mech.}, 948:A23
	
	\bibitem{Samtaney:PF2001}
	Samtaney R, Pullin D~I, and Kosovi{\'c} B
	\newblock 2001
	\newblock {Direct numerical simulation of decaying compressible turbulence and
		shocklet statistics}.
	\newblock {\em Phys. Fluids}, 13:1415--1430
	
	\bibitem{Larkermani:CMAME2024}
	Larkermani E, Bihs H, Winckelmans G, Duponcheel M, Martin T, M{\"u}ller B, and
	Georges L
	\newblock 2024
	\newblock {Development of an accurate central finite-difference scheme with a
		compact stencil for the simulation of unsteady incompressible flows on
		staggered orthogonal grids}.
	\newblock {\em Comput. Methods Appl. Mech. Eng.}, 428:117117
	
	\bibitem{Fornberg:Geopphys1987}
	Fornberg B
	\newblock 1987
	\newblock The pseudospectral method: Comparisons with finite differences for
	the elastic wave equation.
	\newblock {\em Geophysics}, 52:483--501
	
	\bibitem{Verma:SNC2020}
	Verma M~K, Samuel R~J, Chatterjee S, Bhattacharya S, and Asad A
	\newblock 2020.
	\newblock {Challenges in fluid flow simulations using exascale computing}.
	\newblock {\em S.N. Comput. Sci.}, 1:178
	
	\bibitem{Yokokawa:CP2002}
	Yokokawa M, Itakura K, Uno A, and Ishihara T
	\newblock 2002
	\newblock {16.4-Tflops Direct Numerical Simulation of Turbulence by a Fourier
		Spectral Method on the Earth Simulator}.
	\newblock In {\em ACM/IEEE 2002 Conference}, pp. 50--50
	
	\bibitem{Yeung:PNAS2015}
	Yeung P~K, Zhai X~M, and Sreenivasan K~R
	\newblock October 2015.
	\newblock {Extreme events in computational turbulence.}
	\newblock {\em PNAS}, 112:12633
	
	\bibitem{Yang:IEEE2016}
	Yang C, Xue W, Fu H, You H, Wang X, Ao Y, Liu F, Gan L, Xu P, Wang L, Yang G,
	and Zheng W
	\newblock 2016
	\newblock 10m-core scalable fully-implicit solver for nonhydrostatic
	atmospheric dynamics.
	\newblock In {\em SC '16: Proceedings of the International Conference for High
		Performance Computing, Networking, Storage and Analysis}, pp. 57--68
	
	\bibitem{Krasnov:JCP2023}
	Krasnov D, Akhtari A, Zikanov O, and Schumacher J
	\newblock 2023
	\newblock {Tensor-product-Thomas elliptic solver for liquid-metal
		magnetohydrodynamics}.
	\newblock {\em J. Comput. Phys.}, 474:111784
	
	\bibitem{Thibault:CP2009}
	Thibault J and Senocak I
	\newblock 2009
	\newblock {CUDA Implementation of a Navier-Stokes Solver on Multi-GPU Desktop
		Platforms for Incompressible Flows}.
	\newblock In {\em 47th AIAA Aerospace Sciences Meeting including The New
		Horizons Forum and Aerospace Exposition}, pp. 1--15
	
	\bibitem{Czechowski:CP2012}
	Czechowski K, Battaglino C, McClanahan C, Iyer K, Yeung P~K, and Vuduc R
	\newblock 2012
	\newblock {On the communication complexity of 3D FFTs and its implications for
		Exascale}.
	\newblock In {\em Proceedings of the 26th ACM international conference on
		Supercomputing}, pp. 205--214.
	
	\bibitem{Niksiar:IEEE2014}
	Niksiar P, Ashrafizadeh A, Shams M, and Madani A~H
	\newblock 2014
	\newblock {Implementation of a GPU-based CFD Code}.
	\newblock In {\em 2014 International Conference on Computational Science and
		Computational Intelligence}, volume~1, pp. 84--89
	
	\bibitem{Ravikumar:SC2019}
	Ravikumar K, Appelhans D, and Yeung P~K
	\newblock 2019
	\newblock {GPU acceleration of extreme scale pseudo-spectral simulations of
		turbulence using asynchronism}.
	\newblock In {\em Proceedings of the International Conference for High
		Performance Computing, Networking, Storage and Analysis}, SC '19, pp.
	1--22.
	
	\bibitem{Piscaglia:Aerospace2023}
	Piscaglia F and Ghioldi F.
	\newblock 2023
	\newblock {GPU Acceleration of CFD Simulations in OpenFOAM}.
	\newblock {\em Aerospace}, 10:792
	
	\bibitem{Zehner:CF2023}
	Zehner P and Hashimoto A
	\newblock 2023
	\newblock {Acceleration of the data-parallel lower-upper relaxation
		time-integration method on GPU for an unstructured CFD solver}.
	\newblock {\em Comput. Fluids}, 256:105842
	
	\bibitem{Verma:SNC2023}
	Verma M, Chatterjee S, Garg G, Sharma B, Arya N, Kumar S, Saxena A, and Verma
	M~K
	\newblock 2023
	\newblock { Scalable Multi-node Fast Fourier Transform on GPUs}.
	\newblock {\em SN Comp. Sci.}, 4:625
	
	\bibitem{Owens:IEEE2008}
	Owens J~D, Houston M, Luebke D, Green S, Stone J~E, and Phillips J~C
	\newblock 2008
	\newblock {GPU Computing}.
	\newblock {\em Proc. IEEE}, 96:879--899
	
	\bibitem{Li:NCA2019}
	Li F, Ye Y, Tian Z, and Zhang X
	\newblock 2019
	\newblock {CPU versus GPU: which can perform matrix computation
		faster—performance comparison for basic linear algebra subprograms}.
	\newblock {\em Neural Comput. Appl.}, 31:4353--4365
	
	\bibitem{Luebke:IEEE2008}
	Luebke D
	\newblock 2008
	\newblock {CUDA: Scalable parallel programming for high-performance scientific
		computing}.
	\newblock In {\em 2008 5th IEEE International Symposium on Biomedical Imaging:
		From Nano to Macro}, pp. 836--838
	
	\bibitem{Sanders:book:CUDA}
	Sanders J and Kandort E
	\newblock 2010
	\newblock {\em {CUDA by Example: An Introduction to General-Purpose GPU
			Programming}}.
	\newblock Addison-Wesley, Boston
	
	\bibitem{Strontium:book:CUDA}
	Storti D and Yurtoglu M
	\newblock 2016
	\newblock {\em {CUDA for Engineers}}.
	\newblock Pearson Education India, Noida
	
	\bibitem{Ansorge:book:CUDA}
	Ansorge R
	\newblock 2022
	\newblock {\em {Programming in Parallel with CUDA: A Practical Guide}}.
	\newblock Cambridge University Press, Cambridge.
	
	\bibitem{CUDA:Web}
	{NVIDIA Developer}, 2014
	\newblock {CUDA Zone}.
	\newblock [Online; accessed 11-August-2024]
	
	\bibitem{Nishino:CP2017}
	Nishino R and Loomis S H~C
	\newblock 2017
	\newblock {CuPy: A NumPy-compatible library for NVIDIA GPU calculations}.
	\newblock In {\em 31st conference on neural information processing systems},
	volume 151, pp. 16
	
	\bibitem{Cupy:Web}
	{Preferred Networks}, 2020
	\newblock {CuPy: NumPy/SciPy-compatible Array Library for GPU-accelerated
		Computing with Python}.
	\newblock [Online; accessed 9-August-2024]
	
	\bibitem{Cichowlas:PRL2005}
	Cichowlas C, Bona{\"\i}ti P, Debbasch F, and Brachet M~E
	\newblock 2005
	\newblock {Effective Dissipation and Turbulence in Spectrally Truncated Euler
		Flows}.
	\newblock {\em Phys. Rev. Lett.}, 95:264502
	
	\bibitem{Samuel:JOSS2020}
	Samuel R~J, Bhattacharya S, Asad A, Chatterjee S, Verma M~K, Samtaney R, and
	Anwer S~F
	\newblock 2021
	\newblock {SARAS: A general-purpose PDE solver for fluid dynamics}.
	\newblock {\em J. Open Source Softw.}, 6:2095
	
	\bibitem{Harlow:PF1965}
	Harlow F~H and Welch J~E
	\newblock 1965
	\newblock {Numerical calculation of time-dependent viscous incompressible flow
		with free surface}.
	\newblock {\em Phys. Fluids}, 8:2182--2189
	
	\bibitem{Crank:PCPS1947}
	Crank J and Nicolson P
	\newblock 1947
	\newblock {A practical method for numerical evaluation of solutions of partial
		differential equations of the heat conduction type}.
	\newblock {\em Proc. Camb. Phil. Soc.}, 43:50--67
	
	\bibitem{Williamson:JCP1980}
	Williamson J~H
	\newblock 1980
	\newblock {Low-storage Runge-Kutta schemes}.
	\newblock {\em J. Comput. Phys.}, 35:48--56
	
	\bibitem{Gholami:SIAM2016}
	Gholami A, Malhotra D, Sundar H, and Biros G
	\newblock 2016
	\newblock {FFT, FMM, or multigrid? A comparative study of state-of-the-art
		poison solvers for uniform and nonuniform grids in a unit cube}.
	\newblock {\em SIAM J. Sci. Comput.}, 38:C280--306
	
	\bibitem{Briggs:book:Multigrid}
	Briggs W~L, Henson V~E, and McCormick S~F
	\newblock 2000
	\newblock {\em {A Multigrid Tutorial}}.
	\newblock Society for Industrial and Applied Mathematics, Philadelphia, 2nd
	edition
	
	\bibitem{Wesseling:book:Multigrid}
	Wesseling P
	\newblock 1991
	\newblock {\em {An introduction to multigrid methods}}.
	\newblock John Wiley and Sons, West Sussex, 1 edition
	
	\bibitem{Verma:Pramana2013tarang}
	Verma M~K, Chatterjee A~G, Yadav R~K, Paul S, Chandra M, and Samtaney R
	\newblock 2013
	\newblock {Benchmarking and scaling studies of pseudospectral code Tarang for
		turbulence simulations}.
	\newblock {\em Pramana-J. Phys.}, 81:617--629
	
	\bibitem{Strogatz:book}
	Strogatz S~H
	\newblock 2014
	\newblock {\em {Nonlinear Dynamics and Chaos: With Applications to Physics,
			Biology, Chemistry, and Engineering}}.
	\newblock Perseus Books, Reading MA, 2nd edition
	
	\bibitem{Frisch:book}
	Frisch U
	\newblock 1995
	\newblock {\em {Turbulence: The Legacy of A. N. Kolmogorov}}.
	\newblock Cambridge University Press, Cambridge
	
	\bibitem{Lesieur:book:Turbulence}
	Lesieur M
	\newblock 2008
	\newblock {\em {Turbulence in Fluids}}.
	\newblock Springer-Verlag, Dordrecht
	
	
	\bibitem{Pope:book}
	Pope S~B
	\newblock 2000
	\newblock {\em {Turbulent Flows}}.
	\newblock Cambridge University Press, Cambridge
	
	\bibitem{Leslie:book}
	Leslie D~C
	\newblock 1973
	\newblock {\em {Developments in the theory of turbulence}}.
	\newblock Clarendon Press, Oxford
	
	\bibitem{Vincent:JFM1991}
	Vincent A
	\newblock 1991
	\newblock {The spatial structure and statistical properties of homogeneous
		turbulence}.
	\newblock {\em J. Fluid Mech.}, 225:1--20
	
	\bibitem{Hwang:EF2004}
	Hwang W and Eaton J~K
	\newblock 2004
	\newblock {Creating homogeneous and isotropic turbulence without a mean flow}.
	\newblock {\em Exp. Fluids}, 36:444--454
	
	\bibitem{Wilczek:JFM2011}
	Wilczek M, Daitche A, and Friedrich R
	\newblock 2011
	\newblock {On the velocity distribution in homogeneous isotropic turbulence:
		correlations and deviations from Gaussianity}.
	\newblock {\em J. Fluid Mech.}, 676:191--217
	
	\bibitem{Kraichnan:JFM1959}
	Kraichnan R~H
	\newblock 1959
	\newblock {The structure of isotropic turbulence at very high Reynolds
		numbers}.
	\newblock {\em J. Fluid Mech.}, 5:497--543
	
	\bibitem{Dar:PD2001}
	Dar G, Verma M~K, and Eswaran V
	\newblock 2001
	\newblock {Energy transfer in two-dimensional magnetohydrodynamic turbulence:
		formalism and numerical results}.
	\newblock {\em Physica D}, 157:207--225
	
	\bibitem{Kolmogorov:DANS1941Dissipation}
	Kolmogorov A~N
	\newblock 1941
	\newblock {Dissipation of Energy in Locally Isotropic Turbulence}.
	\newblock {\em Dokl Acad Nauk SSSR}, 32:16--18
	
	\bibitem{Orszag:CP1973}
	Orszag S~A
	\newblock 1973
	\newblock {Lectures on the statistical theory of turbulence in fluid dynamics}.
	\newblock In Balian R and Peube J.~L, editors, {\em Les Houches Summer School
		of Theoretical Physics}, pp. 235
	
	\bibitem{Palmer:book:Chaos}
	Palmer K
	\newblock 2000
	\newblock {\em {Shadowing in Dynamical Systems. Theory and Applications}}.
	\newblock Springer, New York
	
\end{thebibliography}
	
	\section*{Nomenclature}
	
\begin{tabular}{ll} 
	FD & Finite-difference \\
	$O$ &  Order of magnitude \\
	MPI & Message Passing Interface \\
	Re & Reynolds number \\
	GPU & Graphics Processing Unit \\
	$\mathbf{u}$ & Velocity field \\
	$t$ & Time \\
	$p$ & Pressure field \\
	$\rho$ &  Density \\
	$\nu$ & Kinematic viscosity \\
	$\mathbf{F}$ & Body force \\
	FMG & Full Multigrid \\
	$u_i$ & $i$th component of the velocity field, $i=\{x,y,z\}$\\
	$\hat{u}_i$ & Fourier transform of the $i$th component of the \\
	& velocity field\\
	$\mathbf{k}$ & Wavenumber vector \\
	$k_i$ & $i$th component of the wavenumber vector \\
	$\hat{p}$ & Fourier transform of the pressure field \\
	$k$ & Magnitude of the wavenumber vector \\
	$u_\mathrm{rms}$ & Root mean square velocity \\
	$L$ & Length of the domain \\	
\end{tabular}

\begin{tabular}{ll}
	$F_0$ & Amplitude of the Taylor-Green force \\
	$\eta$ & Kolmogorov length scale \\
	$k_\mathrm{max}$ & The largest resolved wavenumber \\
	FFT & Fast Fourier Transform \\
	$\epsilon_u$ & Viscous dissipation rate \\
	$E_t$ & Total energy \\
	PDF & Probability distribution function \\
	$\sigma$ & Standard deviation \\
	$\mu$ & Mean \\
	$E(k)$ & Energy spectrum \\
	$\Pi (k)$ & Energy flux \\
	$\mathbf{N}_u(\mathbf{k})$ & Nonlinear term of the Navier Stokes Equation \\
	$\mathbb{R}$ & Real part of a complex number
\end{tabular}
	
	\begin{spacing}{1}

\end{spacing}
\end{document}